\documentclass[aps,prl,twocolumn,superscriptaddresss,showpacs]{revtex4}

\usepackage{amssymb}
\usepackage{amsmath}
\usepackage{colordvi}
\usepackage[colorlinks]{hyperref}
\usepackage{amsthm}
\usepackage{subeqnarray}
\usepackage{cases}
\usepackage{graphicx}
\usepackage{epsfig}
\usepackage{epstopdf}
\usepackage{subfigure}
\usepackage{color}

\def\be{\begin{equation}}
\def\ee{\end{equation}}
\def\bea{\begin{eqnarray}}
\def\eea{\end{eqnarray}}

\makeatletter
\def\NAT@bibsetnum#1{%
 \setlength{\topsep}{\z@}%
 \NATx@bibsetnum{#1}%
}%

\newcommand*{\supplementarystart}{%
  \close@column@grid%
  \clearpage%
  \onecolumngrid%
  \setcounter{enumiv}{0} % resets counter for references
  \setcounter{equation}{0} % resets counter for equations
  \setcounter{figure}{0} % resets counter for figs
  \setcounter{table}{0} % resets counter for tables
  \setcounter{page}{1}
  \c@secnumdepth=4
  \renewcommand{\theequation}{S\arabic{equation}} % equations numbered with S...
  \renewcommand{\bibnumfmt}[1]{[s##1]} % bibtems [S...]
  \renewcommand{\@onlinecite}{s\citealp} % citations [S...]
  \renewcommand{\cite}[1]{{[}\onlinecite{##1}{]}}
  \renewcommand{\thefigure}{s\arabic{figure}}
  \renewcommand{\thetable}{s\Roman{table}}
  \renewcommand{\thepage}{s\arabic{page}}
}

\renewenvironment{widetext@grid}{%
  \par\ignorespaces
  \setbox\widetext@top\vbox{%
   \vskip15\p@
   \hb@xt@\hsize{%
    \leaders\hrule\hfil
    \vrule\@height6\p@
   }%
   \vskip6\p@
  }%
  \setbox\widetext@bot\hb@xt@\hsize{%
    \vrule\@depth6\p@
    \leaders\hrule\hfil
  }%
  \onecolumngrid
  \let\set@footnotewidth\set@footnotewidth@ii
}{%
  \par
  \twocolumngrid\global\@ignoretrue
  \@endpetrue
}%

\makeatother

\begin{document}
\title{Dynamically manipulating topological physics and edge modes in a single degenerate optical cavity}
\author{Xiang-Fa Zhou$^{1,2}$,Xi-Wang Luo$^{1,2}$,Su Wang$^{1,2}$,Guang-Can Guo$^{1,2}$,Xingxiang Zhou$^{1,2}$,Han Pu$^{3,4}$,Zheng-Wei Zhou$^{1,2}$}
\email{zwzhou@ustc.edu.cn$^{1,2}$}
\affiliation{$^1$Key Laboratory of Quantum Information, Chinese Academy of Sciences, University of Science and Technology of China, Hefei, 230026, China \\
$^2$Synergetic Innovation Center of Quantum Information
and Quantum Physics, University of Science and Technology of China, Hefei, 230026, China \\
$^{3}$Department of Physics and Astronomy, and Rice Center for Quantum Materials,
Rice University, Houston, TX 77251, USA \\
$^4$Center for Cold Atom Physics, Chinese Academy of Sciences, Wuhan 430071, P. R. China}
%\date{\today}

%Cavity quantum electrodynamics, 42.50.Pq
%Optical angular momentum (quantum optics), 42.50.Tx
%Topological phases (quantum mechanics), 03.65.Vf
%64.60.Ht 	Dynamic critical phenomena

\begin{abstract}
We propose a scheme to simulate topological physics within a single degenerate cavity, whose modes are mapped to lattice sites.
A crucial ingredient of the scheme is to construct a sharp boundary so that open boundary condition can be
implemented for this effective lattice system. In doing so, the topological properties of the system can manifest themselves on the edge states, which can be probed from the spectrum of output cavity field. We demonstrate this with two examples: a static Su-Schrieffer-Heeger chain and a periodically driven Floquet topological insulator.
Our work opens up new avenues to explore exotic photonic topological phases inside a single optical cavity.
\end{abstract}
\maketitle

{\em Introduction} ---
Exploration of topological physics has become one of the most fascinating frontiers in recent years \cite{nayak2008non,hasan2010colloquium,qi2011topological}. Since Haldane and Raghu proposed to transcribe the topological features of electronic models into photonic ones \cite{haldane2008possible,raghu2008analogs,nayak2008non,hasan2010colloquium,qi2011topological}, studies on topological photonics have been widely developed \cite{joannopoulos2011photonic,wang2008reflection,wang2009observation,hafezi2011robust,fang2012realizing,
khanikaev2013photonic,lu2013weyl,kraus2012topological,rechtsman2013photonic,hafezi2013imaging,
lu2014topological,chen2014experimental,lu2015experimental,schine2016synthetic}.
Although there is no concept of band filling due to the abscence of Pauli exclusion principle, bulk-edge correspondence is still present in this linear bosonic system  \cite{wang2008reflection,wang2009observation,hafezi2011robust,fang2012realizing}.
In such systems, detection of the topologically edge modes are regarded as one of the most important and direct methods of probing their topological properties \cite{hafezi2011robust,khanikaev2013photonic,rechtsman2013photonic,chen2014experimental,lu2015experimental,
liang2013optical,umucalilar2011artificial,skirlo2014multimode,mittal2014topologically,PhysRevX-5-011012,
PhysRevLett.114.037402,PhysRevLett.112.210405}.

Recent studies show that the internal degrees of freedom of quantum systems
\cite{Celi-PRL-112-043001,PhysRevLett.115.195303,PhysRevLett.115.095302,PhysRevA.90.063638,yao2011orbital,price2016synthetic,luo2015quantum}
may be used as synthetic dimensions, which lead to the reduction of physical resources.
In the context of photonics,  it has been shown \cite{luo2015quantum,SM} that synthetic gauge fields in a two-dimensional (2D) system can be effectively simulated by using a 1D array of degenerate cavities \cite{arnaud1969degenerate,collins1970lens,hodgson2005laser}, where the internal degenerate cavity modes serve as an extra dimension.
However, making identical cavities to form the array is extremely challenging in practice.
Hence it is highly desirable that topological phases can be simulated using just a single cavity. Furthermore, how to control the cavity decay and to construct the desired boundary condition for photons are highly nontrivial tasks.

The main purpose of the present work is three-fold: First, we show that it is indeed possible to simulate certain topological phases inside a single cavity. Second, we propose a way to construct a sharp boundary, with which edge states will emerge when the system enters the topological regime.
Finally, exploiting the high controllability of the system, we show how a Floquet topological insulator can be generated by periodically modulating the cavity system. This allows us to investigate Floquet topological phases which possess many unique features not present in static systems \cite{levante1995formalized,asboth2014chiral,dal2015floquet,oka2009photovoltaic,kitagawa2010topological,lindner2011floquet,cayssol2013floquet,gomez2013floquet} and a further understanding of the system \cite{fang2012realizing,rechtsman2013photonic,pasek2014network,asboth2014chiral,dal2015floquet,leykam2016anomalous}.

%%%--------------------------------------------------------------------------------------
\begin{figure}[htpb]
    \centering
    \includegraphics[width=.90\linewidth]{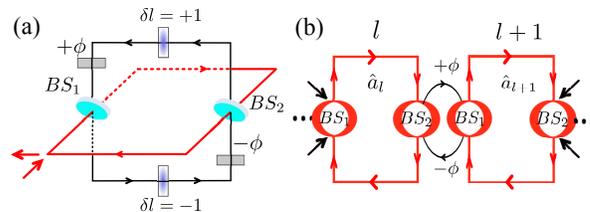}
    \caption{(Color online). (a) Illustration of experiment setup about the degenerate cavity. (b) The effective photonic circuit of (a). $\phi$ is the imbalanced phase between the two arms of the auxiliary cavity.
     }
    \label{fig:1}
\end{figure}
%%%--------------------------------------------------------------------------------------

{\em Effective 1D chain in a single cavity} --- Figure~\ref{fig:1}(a) illustrates schematically the cavity system we will be working with. It contains a main cavity (horizontally oriented in the figure) which supports Laguerre-Gaussian (LG) modes with different orbital angular momenta (OAM), and an auxiliary cavity (vertically oriented in the figure) which is connected to the main cavity by two beam splitters (BS$_1$ and BS$_2$). The electric field of the LG mode, $E^p_l$, is characterized by the radial and the azimuthal quantum numbers $p$ and $l$, respectively. It is possible to make the resonance frequency of the modes to be all degenerate, i.e., independent of $p$ and $l$ (for details, see Ref.~\cite{luo2015quantum,SM}).
For our purpose, the radial quantum number $p$ is irrelevant, and we shall neglect it henceforth.
The azimuthal index $l$ characterizes the OAM of the photon, and the hopping between different OAM modes is accomplished with the aid of the spatial light modulators (SLMs) in the auxiliary cavity, which changes the OAM (i.e., the azimuthal index $l$) of the photon by $\pm \delta l$. Denote the annihilation operator for mode $l$ as $a_l$, different OAM modes are thus mapped to a 1D lattice chain, and the hopping between them is descried by $a^{\dag}_{l} a_{l+\delta l} + h.c.$ in the lattice model.
In general, arbitrary long-range hopping can be realized by adjusting $\delta l$. The hopping strength is determined by the transmission/reflection coefficients of the beam splitters and can be further controlled by the phase retarders $\pm \phi$ \cite{SM}. The effective lattice system with nearest-neighbor hopping is schematically illustrated in Fig.~\ref{fig:1}. Note that the effective lattice size can be doubled if we take into account that a given $l$-mode comes with two orthogonal polarizations (see below).

{\em Creating sharp boundary} --- To realize open boundaries for the effective lattice system, we would like the cavity to have an OAM cutoff $L_m$, such that modes with $l=0$, 1, ..., $L_m$ have negligible decay rates whereas all other modes are not supported. The $l=0$ LG mode is the usual Gaussian mode with intensity peaks at the center, whereas $l\neq 0$ LG modes all have doughnut shape whose intensities peak at a circle with radius scaled as $\sqrt{l}$. This spatial intensity distribution and the finite size of the cavity mirrors may lead to $l$-dependent decay rate.  Such soft boundary due to the $\sqrt{l}$ scaling can cause serious loss of photons with large $l$ \cite{SM} and destroy all interesting physics related to edge modes (see Fig. \ref{fig:4}(c)).

\begin{figure}[htpb]
    \centering
    \includegraphics[width=.90\linewidth]{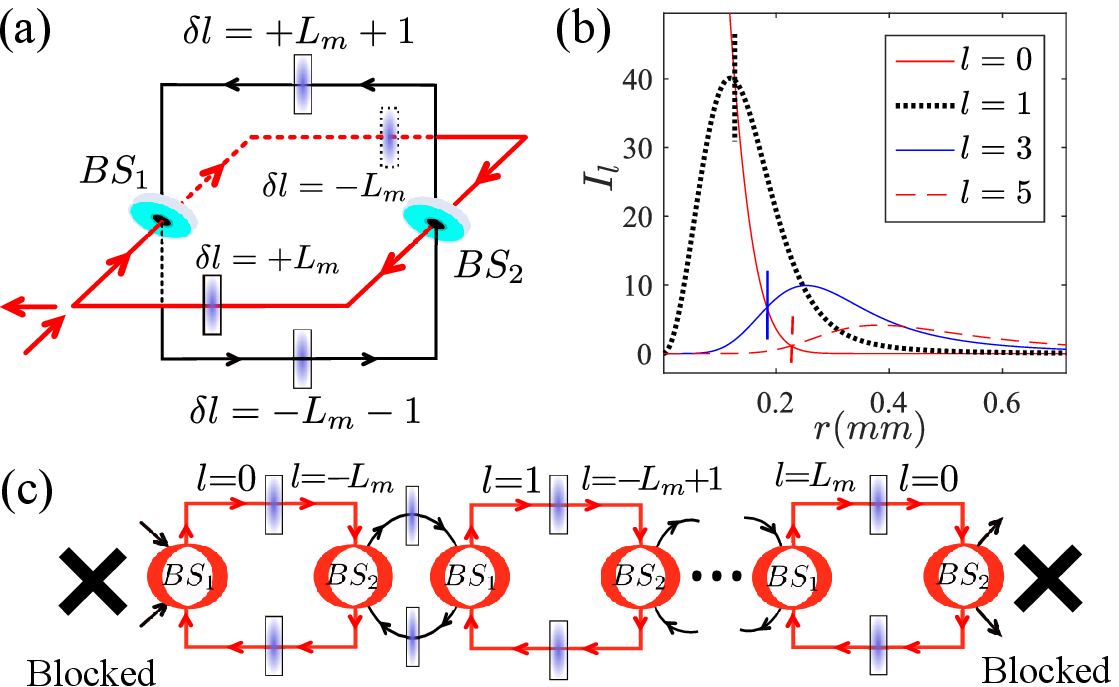}
    \caption{(Color online) (a) Proposed experimental setup of single degenerate cavity to simulate 1D finite lattice using composite modes induced by two SLMs with $\delta l = \pm L_m$. Two hollow beam splitters are employed so that only modes $l=0$ can transmit through the hole. (c) is the effective photonic circuit of (a).  (b) Normalized intensity profiles $I_l(r)=|E_l(r)|^2$ for different $l$-modes calculated using the parameters discussed in \cite{SM}. The vertical lines indicate the possible center pinhole sizes in each BS for different hopping steps $n=1$, 3, and 5, with the corresponding fraction of $l=0$ photon intensity inside the pinhole as 78\%, 96\%, and 99\%, respectively. }
    \label{fig:2}
\end{figure}

To create a sharp boundary, we take advantage of the fact that the $l=0$ mode can be easily distinguished from the $l \neq 0$ modes and modify the cavity system as schematically shown in Fig.~\ref{fig:2}(a). Here we put two SLMs in the main cavity with $\delta l= \pm L_m$, respectively. For photons traveling in the main cavity in the direction as shown by the arrows, their OAM will change when passing the SLMs. Specifically, a photon with azimuthal index $l$ to the left of the SLMs will change it to $l-L_m$ when traveling to the right of the SLMs. Hence the mode in the main cavity becomes composite and we label this mode pair $[l, l-L_m]$ as $|l \rangle$. We make the  two SLMs in the auxiliary cavity to have $\delta l = \pm (L_m+1)$, respectively. Finally a hole is made in each of the two beam splitters connecting the main and the auxiliary cavities. The hole size is carefully designed so that, ideally, a photon with $l=0$ will always go through the hole without being reflected, while all other modes with $l \neq 0$ will be reflected with certain probability. It is not difficult to see that \cite{SM}, with this design, (1) a composite mode $|0 \rangle$ in the main cavity may hop to $|1 \rangle$, but not to $|-1 \rangle$; (2) a composite mode $|L_m \rangle$ may hop to $|L_m-1 \rangle$, but not to $|L_m +1 \rangle$; (3) a composite mode $| l\rangle$ with $0<l<L_m$ may hop to either $|l-1 \rangle$ or $|l+1 \rangle$. In other words, we have succeeded in creating two sharp boundaries such that only composite modes $|l=0, 1, \cdots, L_m \rangle$ can exist in the main cavity.

%%%--------------------------------------------------------------------------------------
%%%--------------------------------------------------------------------------------------

It is clear that the key here is the design of the hole in the beam splitters which should let $l=0$ mode pass through with high probability, while not affecting too much the $l \neq 0$ modes. The sharpness of the boundary is then determined by how well we can distinguish $l=0$ photon from $l \neq 0$ modes. We can achieve good distinguishability due to the small intensity overlap between $l=0$ mode with the adjacent $l=\pm 1$ modes.
Experimentally, sharper boundaries can be obtained for larger hopping step $n$ if we replace the two SLMs in the auxiliary cavity
with $\delta l= \pm (L_m+n)$. In this way, the effective lattice sites are represented by modes $|l=0, n, \cdots, nL_m \rangle$. We need only to distinguish $l=0$ mode from $l= \pm n$ modes whose intensity overlap scales as $e^{-n}$ (see Fig.~\ref{fig:2}(b) and details in \cite{SM}).

With the creation of sharp boundaries, we can now use it to explore the topological properties of the system. We will use two examples below to demonstrate this.

%%%--------------------------------------------------------------------------------------
\begin{figure}[htpb]
	\centering
	\includegraphics[width=0.95\linewidth]{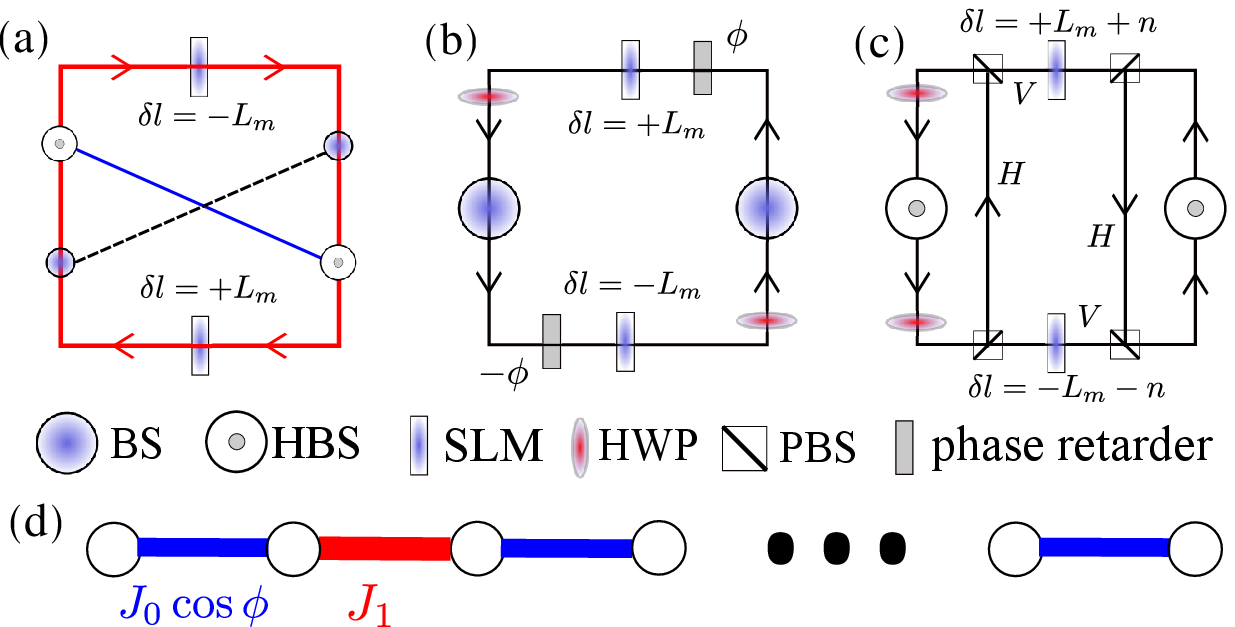}
	\caption{(Color online) Schematic diagram of simulating 1D SSH chain. (a) shows the skeleton inside of the main cavity with two-auxiliary-cavity circuits depicted in (b) and (c), where optical circuits related to the hopping $J_0 \cos\phi$ and $J_1$ are shown. (d) is the diagrammatic representation of the Hamiltonian $H$. BS: beam splitter. HBS: BS with a pinhole in the center. SLM: spatial light modulator. HWP: half-wave plate. PBS: polarization beam splitter which transmit vertical polarized photons and reflect horizontal polarized photons.   }
	\label{fig:3}
\end{figure}
%%%--------------------------------------------------------------------------------------

\begin{widetext}

%%%--------------------------------------------------------------------------------------
\begin{figure}[htpb]
    \centering
    \includegraphics[width=0.95\linewidth]{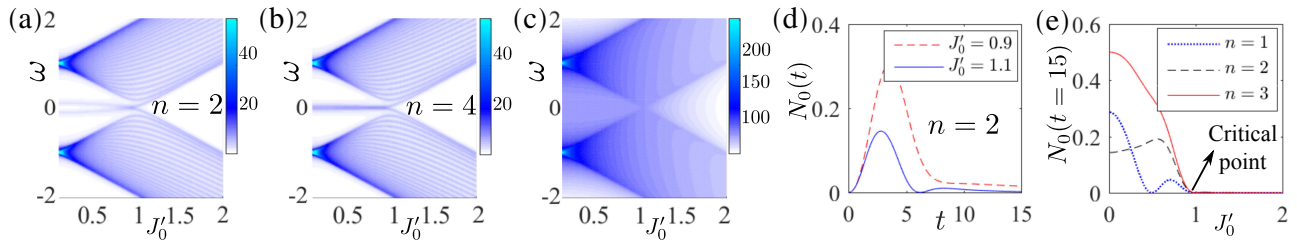}
    \caption{(Color online) (a) and (b) show the spectrum $\tau(\omega)$ of model (\ref{Hamilton1}) for $J_1=1$ and $L_m=49$ with the decay rates $\gamma_j=0.05(1+e^{-j/\sqrt{25}}+e^{-|j-L_m|/\sqrt{25}})$ for $n=2$ and $n=4$ respectively, where the influence of $H'$ in remaining lattices is also taken into account. (c) shows the output spectrum of a prolonged chain using the same decay as in (a), which corresponds to the soft boundary condition discussed in the context. (d) shows the dynamics of $N_0$ with hopping step $n=2$ for an input pulse $a(t)_{in,0}=\exp[-(t-3)^2/8]/\sqrt{2\sqrt{\pi}}$ at site $j=0$ with the decay given in (a) for fixed
$J_0'=0.9$ and $1.1$ respectively. (e) plots the amplitude $N_0$ for different $n$ at $t=15$ along with $J_0'$, which drops to zero across the phase transition point. }
    \label{fig:4}
\end{figure}
%%%--------------------------------------------------------------------------------------
\end{widetext}

{\em Realizing and probing SSH model} --- Our first example concerns the realization of the Su-Schrieffer-Heeger (SSH) model \cite{su1980soliton}, a prototypical 1D topological model, as schematically illustrated in
Fig.~\ref{fig:3}. Here we take advantage of the fact that each OAM mode $l$ comes with two orthogonal polarizations, which we will map to lattice sites as $E_{l,H} \rightarrow 2l$ and $E_{l,V} \rightarrow2l+1$.
The coupling between modes $a_{2l}$ and $ a_{2l+1}$ can be easily accomplished with polarization rotators inside the auxiliary cavities.
The coupling $a_{2l+1} \leftrightarrow a_{2l+2}$ corresponds to a polarization-dependent hopping $E_{l,V} \leftrightarrow E_{l+1,H}$, which can be realized with the combination of the polarized beam splitters (PBSs) and the SLMs (See Fig.3(c) and \cite{SM} for details). The total effective Hamiltonian simulated can then be written as $H_T=H+H'$ with
\bea
H = \sum_{l=0}^{L_m} \big[J_0\cos(\phi) a^{\dag}_{2l} a_{2l+1} +J_1 \alpha^{(n)}_{l+1} a^{\dag}_{2l+1} a_{2l+2} + h.c.\big]\,,
\label{Hamilton1}
\eea
where $H$ is the desired SSH model Hamiltonian when $\alpha^{(n)}_{l+1}=1$.
$H'$ describes the interaction of other cavity modes in the remaining lattice sites, whose explicit form can be found in \cite{SM}.
The phase dependent coupling proportional to $\cos\phi$ is due to the interference effect inside the auxiliary cavity, which can be used as a convenient control knob to adjust the hopping amplitude $J'_0 \equiv J_0 \cos \phi$ \cite{SM}.
The presence of pinhole results in a reduction of coupling strength at the lattice site $2(l+1)$ defined by $\alpha^{(n)}_{l+1}=1-\eta^{(n)}_{l+1}$ for giving hopping step $n$, where $\eta^{(n)}_{l+1}$ is the portion of photons for modes $|(l+1)n\rangle$ inside the pinhole of the BSs.
As shown in \cite{SM}, $\eta^{(n)}_{l+1}$ decreases exponentially along with $l$.
In the ideal case $\alpha^{(n)}_{l+1}=1$, the system becomes topologically nontrivial when $J'_0<J_1$. In the presence of sharp boundaries, this leads to topologically induced edge states.

To illustrate how the presence of the edge states can be detected, let us calculate the output spectrum of the cavity using the Langevin equations \cite{walls2007quantum,luo2015quantum}
\bea
\partial_t a_j=-i[a_j,H(t)]-\frac{\gamma_j}{2} a_j -\sqrt{\gamma_j}a_{in,j}\,,
\eea
with $\gamma_j$ the decay rate on lattice site $j$. The output field is linked to the dynamics inside the cavity through the standard input-output relation
$a_{out,j}(t)=a_{in,j}(t)+\sqrt{\gamma_j}a_j(t)$. In the frequency domain, this leads to
$a_{out,j}(\omega)=\sum_{j'} (\delta_{jj'}-T_{jj'})a_{in,j'}(\omega)$ with the transmission element $T_{jj'}=-i\langle j| \sqrt{\Gamma}[\omega-H+i\Gamma/2]^{-1}\sqrt{\Gamma}|j'\rangle$ and the decay matrix $\Gamma=\mbox{diag}\{\gamma_0,\cdots,\gamma_{Lm}\}$.

Figure \ref{fig:4}(a) and (b) show the total transmission rate $\tau(\omega)=\sum_{j,j'}|T_{jj'}|^2$ as functions of $J'_0$ for $n=2$ and $4$ respectively. The imperfection induced by the pinhole in the BSs results in site-dependent hoppings characterised by $\alpha^{(n)}_{l+1}$ and unwanted coupling $|0\rangle \rightarrow |-n\rangle$ and $|L_m\rangle \rightarrow |L_m+n\rangle$ at boundaries \cite{SM}, both of which are  explicitly taken into account. For $n=2$, the presence of such unwanted tunneling couples bilateral edge modes  in the topological nontrivial regime.
This results in the splitting of edge modes into two branches around $\omega=0$ \cite{SM}. For larger hopping step $n=4$, the two branches merge as such hopping decreases exponentially with $n$.
For comparison, we plot the transmission rate for a soft boundary in Fig.~\ref{fig:4}(c), where the presence of edge modes is completely erased. This clearly demonstrates the importance of constructing the sharp boundary in our system.
Figure~\ref{fig:4}(d) illustrates the dynamics of $N_0=\langle a^{\dag}_0a_0 \rangle$, the amplitude of the first site for $n=2$, when initially we inject an input pulse
from this site. In the topological regime $J_0'/J_1<1$, due to the presence of the edge state, $N_0$ persists
over a very long time; whereas in the nontopological regime when $J_0'/J_1>1$, $N_0$ decays to zero rather quickly. In Fig.~\ref{fig:4}(e), we plot the value of $N_0$ at $t=15$ as a function of $J'_0$ when $J_1$ is fix to be 1 for different $n$. A transition at $J_0'/J_1=1$ can be easily seen, which can be viewed as a clear evidence of the topological phase transition at that critical point \cite{SM}. We note that the oscillation shown for small hopping step is due to the interference of two split edge modes in the presence of unwanted hopping at boundaries. For larger $n$, such oscillatory behavior disappears.

{\em Floquet topological insulator and edge modes inside cavity} --- In the second example, we take advantage of the
flexibility of our cavity system and also investigate a periodically driven situation.
Such Floquet systems have received great attention recently as they exhibit many unique properties absent in the static models \cite{oka2009photovoltaic,kitagawa2010topological,lindner2011floquet,cayssol2013floquet,gomez2013floquet}. We periodically modulate the system by modulating the phase delay as $\phi(t)=\phi_0+\alpha\cos(\Omega t/2)$, which leads to a periodic modulation of the hopping amplitude $J'_0=J_0 \cos \phi$.
When $\phi_0=0$, we have $\cos[\alpha\sin(\Omega t/2)]=j_0(\alpha)+2j_2(\alpha)\cos(\Omega t) + ...$ with $j_n(x)$ the $n$th order Bessel function.
We can choose $\alpha$ such that all high-order terms can be safely neglected. This leads to a Floquet version of Hamiltonian (\ref{Hamilton1}) where the hopping $J_0 \cos \phi$ is now replaced with the modified one as $J_0+\lambda\cos(2\pi t/T)$ with $T=2\pi/\Omega$. Experimentally, such high frequency phase modulation of $\phi$ can be implemented with the aid of an electro-optic modulator, where the modulation frequency can be as high as tens of GHz.
This is much larger than the typical cavity coupling strength (a few MHz), and is sufficient for our purpose.

The properties of such periodic driven system can be obtained using the standard Floquet theory \cite{asboth2014chiral,dal2015floquet,pasek2014network,leykam2016anomalous,levante1995formalized,SM}.
The quasi-energies $\epsilon_q$ and Floquet modes can be solved in the composite Floquet space $T \otimes R$ where $R$ represents the usual Hilbert space and $T$ is spanned by the periodic functions $\langle t|m\rangle=e^{im\Omega t}$ \cite{SM}. The index $m$ describes the number of phonons and defines the subspace named as the $m$th Floquet replica.
The wave function in the usual Hilbert space $|\psi\rangle=\sum_{m,j}c_{m,j}(t)\exp(im\Omega t)|j\rangle$ can be rewritten as $|\psi^F\rangle=\sum_{m,j}c_{m,j}(t)|m,j\rangle$, which satisfies the modified Schr\"{o}dinger equation \cite{levante1995formalized}
\bea
\frac{d|\psi^F\rangle}{dt}=-i \mathbb{H}_F |\psi^F\rangle - \frac{\gamma}{2}|\psi^F\rangle\,,
\eea
with the last term describing the dissipation effect. The time-independent Floquet Hamiltonian
reads $\mathbb{H}_F=\hat{F}_m\otimes\hat{H}^{(m)}+\Omega \hat{F}_z\otimes I_R$,
where $I_R$ is the identity operator in $R$-space, $H(t)=\sum_m \hat{H}^{(m)}e^{im\Omega t}$, $(F_m)_{i,j}=\delta_{j,i+m}$, and $(F_z)_{m,n}=m\Omega\delta_{m,n}$.

For high driving frequency $\Omega$, different Floquet bands are almost uncoupled. When $\Omega$ decreases, the interaction of Floquet replicas for $m=0$ and $m=\pm 1$ leads to the appearance of gaps at $\pm \Omega/2$ . Topological transition occurs when bands in different replicas start to overlap with each other, and is signalled by the presence of edge states at quasi-energy $\epsilon_q=0$ and $ \Omega/2$, respectively.

%%%--------------------------------------------------------------------------------------
\begin{figure}[htpb]
    \centering
    \includegraphics[width=0.950\linewidth]{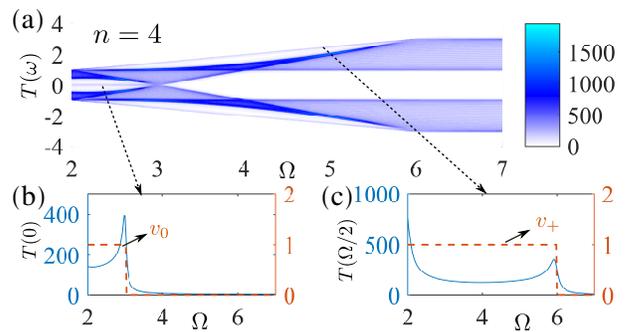}
    \caption{(Color online). (a) The spectrum $T(\omega)$ within the Floquet zone $(-\Omega/2, \Omega/2)$ for different $\Omega$ for hopping step $n=4$ with the same decay used in Fig. \ref{fig:4}. Other parameters are $[J_0,J_1,\lambda]=[2,1,0,1.6]$ and $L_m=49$. (b) and (c) show the spectra $T(0)$ and $T(\Omega/2)$ along with $\Omega$, where topological transition is manifested by jumps around the phase transition points. $v_0$ and $v_+$ are their corresponding numbers of edge modes defined in \cite{SM}.
  }
    \label{fig:5}
\end{figure}
%%%--------------------------------------------------------------------------------------

As in the previous example, the presence of the Floquet phase transition and the associated edge states can be observed by detecting the total output spectrum defined as $ T(\omega)=\sum_{\psi^F(0)}\sum_{m,j} |c_{m,j}(\omega)|^2$
for $\omega \in (-\Omega/2,\Omega/2)$ \cite{SM}.
The input state $\psi^F(0)$ can be prepared by feeding the cavity using mode $j$ with different frequency $\omega=m\Omega$.
When $\omega$ is resonant with the Floquet modes, it induces a peak in the spectrum.
Especially around $\omega=0$ or $\Omega/2$, $T(\omega)$ is almost completely determined by the
presence of the mid-gapped edge modes, while contributions from other modes are greatly reduced.
This provides a direct evidence of the Floquet topological phase transitions.

Figure \ref{fig:5} shows the cavity output spectrum as a function of $\Omega$ within the Floquet zone, where the size effect of pinhole in the BSs is also involved.
The presence of the Floquet gaps is revealed by the vanishing $T(\omega)$ around quasi-energy $0$ and $\Omega/2$.
In addition, starting with a topologically trivial phase at large $\Omega$, Floquet topological phase transitions occur when two replicas touch each other as $\Omega$ decreases.
The construction of boundaries enable us to detect such transitions by observing the output spectrum directly. As shown in Fig.~\ref{fig:5}(b) and (c), due to the presence of finite gaps, the amplitude of $T(0)$ and $T(\Omega/2)$ exhibit staircase-like structure and jump around the critical point where the phase transition occurs.
This can be viewed as a direct evidence of Floquet topological phase transitions.

{\em Outlook and Conclusion} ---
We have proposed a scheme to simulate topological physics in a single optical cavity by constructing sharp boundaries in the synthetic dimensions.
All the operations about the photonic OAM modes proposed here can be reliably implemented through linear elements, which make the system experimental friendly and resource undemanding.
The proposed scheme can also be extended to explore nontrivial topological physics in high dimensional system \cite{harper1955,Aubry1980}.
In view of current experimental progress on synthetic magnetic field for photons \cite{schine2016synthetic} and the strong light-atom coupling inside a multimode resonator \cite{kollar2016supermode},
effective photon-photon interactions in degenerate cavity regime \cite{schmidt1996giant,imamoḡlu1997strongly,harris1998photon} also becomes possible. Therefore our work also opens up an avenue to explore various exotic topological photonic states in optical cavity system.

{\em Acknowledgement} ---
XFZ thanks Jin-Shi Xu for helpful discussions. The authors thank the anonymous referees for many valuable suggestions. This work was funded by National Natural Science Foundation of China (Grant Nos. 11574294, 61490711, 11474266), the Major Research plan of the NSFC (Grant No. 91536219), the National
Key Research and Development Program (Grant No. 2016YFA0301700),and the "Strategic Priority Research Program(B)" of the Chinese Academy of Sciences (Grant No. XDB01030200). HP is supported by the US NSF (Grant No. PHY-1505590) and the Welch Foundation (Grant No.
C-1669).

\bibliographystyle{apsrev4-1}

\bibliography{tex-total}

%%%%%%%%%%%%%%%%%%%%%%%%%%%%%%%%%%%%%%%%%%%%%%%%%%%%%%%%%%%%%%%%%%%%%%%%%%%%%%%%%55
%%%%%%%%%%%%%%%%%%%%%%%%%%%%%%%%%%%%%%%%%%%%%%%%%%%%%%%%%%%%%%%%%%%%%%%%%%%%%%%%%%%

%%%%%%%%%%%%%%%%%%%%%%%%%%%%%%%%%%%%%%%%%%%%%%%%%%%%%%%%%%%%%%%%%%%%%%%%%%%%%%%%%%%%%
%%%%%%%%%%%%%%%%%%%%%%%%%%%%%%%%%%%%%%%%%%%%%%%%%%%%%%%%%%%%%%%%%%%%%%%%%%%%%%%%%%%%%

\end{document}

%================Supplementary===============
\supplementarystart

\centerline{\bfseries\large SUPPLEMENTAL MATERIAL}
\vspace{6pt}

\centerline{\bfseries\large Dynamically manipulating topological physics and edge modes in a single}
\vspace{6pt}
\centerline{\bfseries\large  degenerate optical cavity}
\vspace{6pt}
\centerline{Xiang-Fa Zhou, Xi-Wang Luo, Su Wang, Guang-Can Guo, Xingxiang Zhou, Han Pu, Zheng-Wei Zhou}

\maketitle

\vspace{.5cm}

In this Supplemental Material, we provide some basic background about degenerated cavity where technique details inside the ancillary circuits are clarified to implement the desired hoppings. The reduction of hopping amplitudes due to the size effect of pinhole in each beam splitter is also discussed. The calculation details about Floquet topological system are also provided.

\section{Degenerated cavity and effective construction of 1D lattice with boundaries}

\begin{widetext}
For a ring-type cavity which is made of optical elements with cylindrical symmetry, the cavity mode are the
Laguerre-Gaussian (LG) modes $E_{p,l}$, with $p,l$ the radial and azimuthal index respectively. Its resonance frequency is determined by
\begin{equation}
kL_0 - (2p+l+1)\arccos \frac{A+D}{2} = 2n\pi,
\label{eq:mode_freq}
\end{equation}
where $L_0$ is the length of the round-trip optical path,
and $A$ and $D$ are diagonal elements in the round-trip ray matrix. The
off-diagonal elements of the round-trip ray matrix, $B$ and $C$,
only affect the beam waist of the resonance modes. $n$ is an integer.
If we design the cavity such that $A=D=1$ and $B=C=0$, then the resonance frequency
becomes independent of $p$ and $l$, and such a
cavity is called a degenerate cavity \cite{luo2015quantum,arnaud1969degenerate}. It can support photon modes of
different $p$ and $l$ simultaneously.
The LG mode $E_{l,p}$ carries an OAM of $l\hbar$ per photon, thus the degenerate cavity could support photon modes
in different OAM states. Our simulator are shown in Fig. \ref{fig:cavity1}, and the optical designs are:

\begin{enumerate}
 \item The length of the main cavity is chosen for constructive
interference, $kL_0=2n\pi$. The elements of the ray matrix
for the optical paths $BS_1\rightarrow BS_2$ and $BS_2\rightarrow
BS_1$ in the main cavity are $A=D=-1$, $B=C=0$.

 \item The auxiliary cavity consisting of the two beam splitters $BS_1$,
$BS_2$ and the two spatial light modulators $SLM_1$, $SLM_2$. Its length
is chosen for destructive interference, $kL_0=(2n+1)\pi$. The elements of the
ray matrix for optical paths $SLM_1\rightarrow BS_2\rightarrow SLM_2$ and
$SLM_2\rightarrow BS_1\rightarrow SLM_1$ are $A=D=-1$,
$B=C=0$.
\end{enumerate}

%%%--------------------------------------------------------------------------------------
\begin{figure}[htpb]
    \centering
    \includegraphics[width=0.70\linewidth]{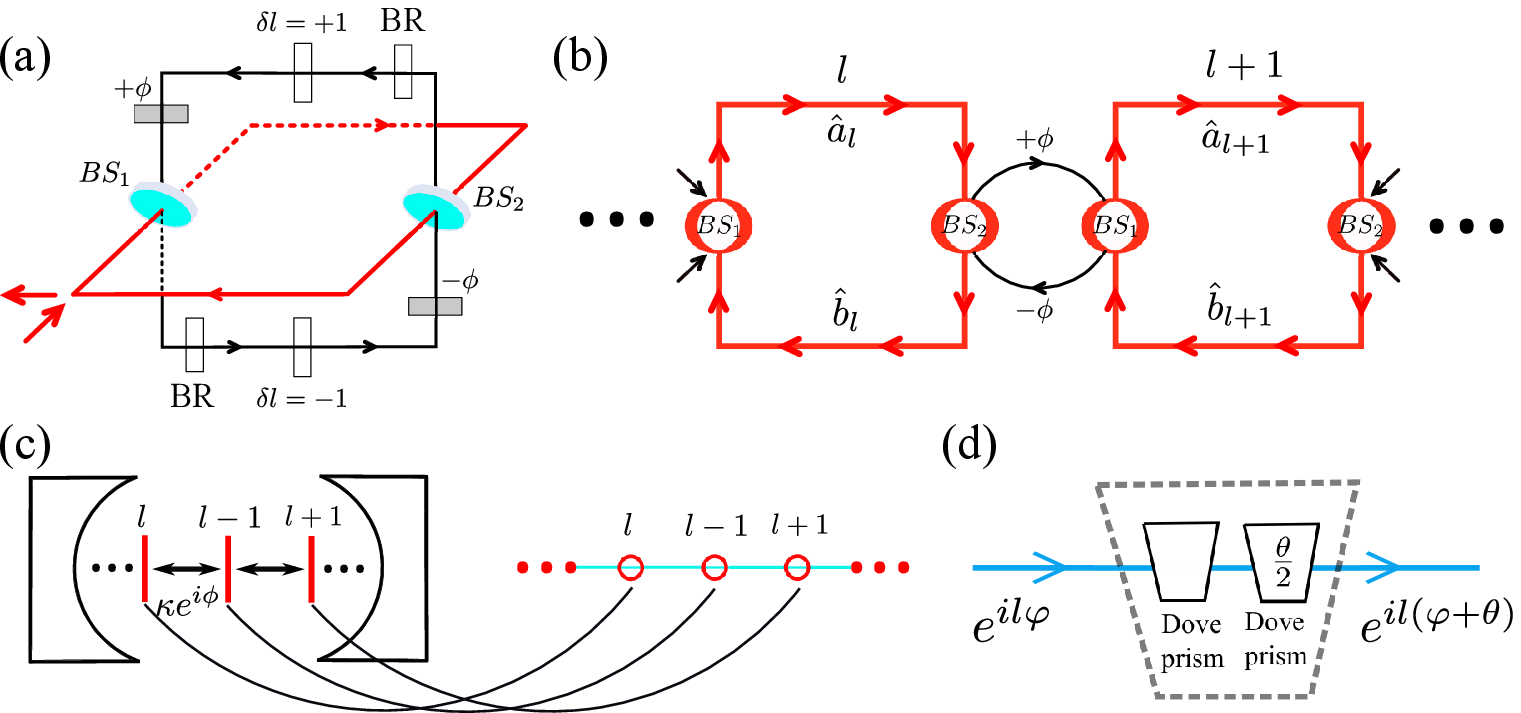}
    \caption{(Color online).  (a) Illusion of experiment setup about the degenerate cavity of our simulator. (b) is the effective photonic circuit of (a). $\phi$ is the imbalanced phase between the two arms of the auxiliary cavity. (c) Mapping of the simulator to a 1D lattice. BR is beam rotator that rotates the light beam by an angle of $\theta$ as shown in (d). It introduces a OAM-dependent phase to the beam $e^{il\varphi}\rightarrow e^{il(\varphi+\theta)}$.  }        \label{fig:cavity1}
\end{figure}
%%%--------------------------------------------------------------------------------------

\end{widetext}

Such a simulator is conceptually equivalent to a 1D lattice, with the lattice sites represented
by the OAM states (See Fig. \ref{fig:cavity1}). The SLMs change the OAM of light by $\delta_l=\pm1$ depending on the incident direction.
The BSs split a portion of the light in the main cavity to the SLMs and merge it back, this corresponds to photon
tunneling between neighbor lattice sites. In addition, long-range tunneling could be realized by a separate auxiliary cavity consisting SLMs that change the OAM of photon by $\pm m$ with $m>1$.
The Hamiltonian of the system reads
\begin{eqnarray}\label{eq:tight-bind}
\mathcal{H}=&-&\kappa\sum_{l}\left(e^{i\phi}\hat{a}_{l+1}^{\dagger}\hat{a}_{l}+h.c.\right),
\end{eqnarray}
where the tunneling strength is $\kappa=\Omega_0 |r|^2 /4 \pi$ with $r$ the reflectivity coefficient and
$\Omega_0$ the free spectral range of the main cavity. The tunneling phase $\phi$ is determined by the phase imbalance between the two arms of the auxiliary cavity, it could be either OAM independent or OAM dependent.
A beam rotator, consisting of two Dove prisms which are rotated by $\theta/2$ with respect to each other, will rotate the beam by an angle of $\theta$, based on which, we obtain an OAM dependent phase $e^{il\theta}$ to the light beam, as shown in Fig. \ref{fig:cavity1}.
Putting a beam rotator with rotating angle $\pm\theta$ in each arm of the auxiliary cavity, we obtain a OAM dependent tunneling phase $\phi=\phi_0+l\theta_0$, with $\phi_0$ the phase imbalance caused by the optical path length.

The tunneling coefficient could be tuned by an interference mechanism. If we introduce another auxiliary cavity with tunneling phase $\phi=\phi_1+l\theta_1$, then the tunneling coefficient would become $\kappa [e^{i(\phi_0+l\theta_0)}+e^{i(\phi_1+l\theta_1)}]$. In particular, if $\phi_0=-\phi_1$ and $\theta_0=-\theta_1$, the tunneling coefficient becomes $\kappa \cos(\phi_0+l\theta_0)$.

\begin{figure}[htpb]
    \centering
    \includegraphics[width=.70\linewidth]{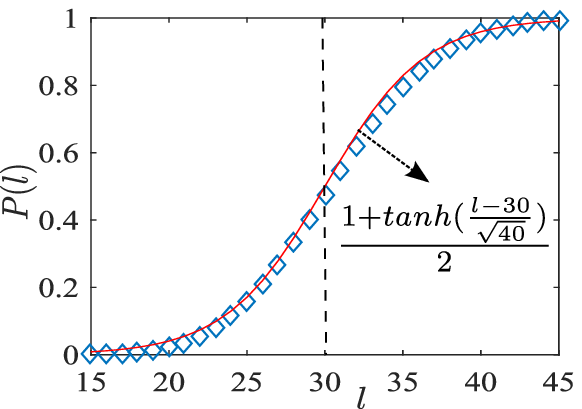}
    \caption{(Color online). The portion $P(l)$ of photons outside the cavity with fixed mirror size $r=\sqrt{30/2}$, which increases exponentially as $l\rightarrow 30$. The data is fitted by the red solid line.
     }
    \label{fig:s2}
\end{figure}

When the beam sizes is comparable with the cavity mirror, leakage of photons outside the cavity is unavoidable due to the typical intensity distributions of LG modes.
For instance, the intensity profile of $E^{p=0}_l$ mode reaches its maximal when $r=\sqrt{l/2}$.
We stress that for $p \neq 0$, the mode becomes more spatially extended along the radial direction.
In experiments, the cavity modes can be excited by feeding the cavity with Gaussian beams $E^0_0$ such that the waists of the beams overlap with the SLMs. Using this setting, all high order modes with $p>0$ can be greatly suppressed.
When $l\gg1$, two adjacent OAM modes $E^0_l$ and $E^0_{l+1}$ becomes almost completely overlap with each other, which makes it extremely difficult to distinguish them from their intensity distributions.
In addition, if we consider a finite radius $\sqrt{L/2}$ of cavity mirrors, the portion of photons $P(l)=\int_{\sqrt{L/2}}^{\infty}dr r |E^0_l|^2$ outside the radius increases almost exponentially as $l \rightarrow L$ (see Fig. \ref{fig:s2}).
This leads to the fast increasement of decay rate for cavity modes near the nature boundaries of the effective lattice. For topological nontrivial system, these factors result in serious photonic loss which can annihilate almost all interesting phenomena related to edge physics (see Fig. 4 in the paper).

%%%--------------------------------------------------------------------------------------
%%%--------------------------------------------------------------------------------------

To construct an effective boundaries of the 1D chain, we make use of the fact that only the zero OAM mode has a high intensity at the beam center, in contrast with doughnut beam shapes for $l\neq 0$ modes. We make proper design of the optical circuit so that we select $l=0$ as boundary by spatial filtering.
This is achieved by introducing composite structure of the cavity modes with the help of hollow BSs and SLMs. The explicit construction is depicted in Fig. 2. Specifically, if we start with a composite mode $[0, -L_m]$ inside the main cavity,
this modified auxiliary cavity can only induce an effective hopping between modes $[0,-L_m] \rightarrow [1,-L_m+1]$, where photons can only be reflected into the auxiliary circuit through $BS_2$ on the right arm in the main cavity.
The SLMs with $\delta l=\pm(L_m+1)$ in the auxiliary circuit induces a photonic OAM changing from $l=-L_m+1$ to $l=1$, which ensures that photons can then be diverted back to the main cavity through $BS_1$ and overlap with the modes $[1,-L_m+1]$.
We note that the tunneling $[0,L_m] \leftrightarrow [-1,-L_m-1]$ is blocked since
a hollow $BS_1$ on the left arm cannot reflect photons with $l=0$ into the auxiliary circuits.
Similar discussion also applies to the mode $[L_m,0]$.
Therefore, using this special setting, one can realize an effective finite hopping chain as $[0,-L_m]\leftrightarrow[1,-L_m+1] \cdots [L_m,0]$ with $[0,-L_m]$ and $[L_m,0]$
acting as the effective boundaries.

As an example, in the main text, we have shown the explicit construction of a finite SSH chain based on the above design, where two different ancillary circuits are introduced to implement the corresponding two hopping terms shown in Eq. (1). Here for given OAM $l$ of photon mode $E_l$, we map it to two different lattice sites labeled as $2l$ and $2l+1$ by taking into account their inner polarization degrees of freedom $a_{2l} \rightarrow E_{l,H}$ and $a_{2 l+1} \rightarrow E_{l,V}$
with $H$ and $V$ corresponding to the horizontal and vertical polarized states respectively.
The hopping $a^{\dag}_{2l}a_{2l+1}+h.c.$ involves only cavity modes with the same OAM $l$ and different polarizations, which can be implemented by introducing spin-flip operator inside the auxiliary cavity, as shown in Fig.`(3b) in the main text. Since the upper and lower arms in the auxiliary cavity depicted in Fig.~3(b) provide two different polarization flip channels with opposite phase retardation, the interference of the two arms results in a phase dependent coupling proportional to $\cos\phi$, and as a result the phase delay $\phi$ becomes a convenient control knob to adjust the hopping amplitude $J'_0 \equiv J_0 \cos \phi$.

The hopping Hamiltonian $a^{\dag}_{2l+1}a_{2l+2}+h.c.$ corresponds to  the coupling of optical modes $E_{l,V} \leftrightarrow E_{l+1,H}$, which becomes polarization-dependent as only OAM of vertical (horizontal) polarized modes can be increased (decreased) through the ancillary circuit. To implement such polarization-dependent OAM hopping $l \rightarrow l+1$, we introduce two additional optical circuits for horizontal polarized modes with the help of polarized beam-splitters (PBS) (See Fig. 3c). The PBSs are designed such that only vertical polarized photons can go through while horizontal polarized beams are reflected.  One can see that for an incoming optical mode $E_{l-Lm}$ from the right BS, only vertical polarized mode $E_{l-L_m,V}$ can pass through the SLM at the top of the circuit, which realizes the hopping $E_{l-L_m,V}\rightarrow E_{l+1,V}$. The polarization of this mode is then flipped by inserting a HWP before it reaches the left BS and couples back into the main circuit. We note that the horizontal mode $E_{l-L_m,H}$ has been reflected back to the right BS by PBSs to hinder the unwanted hopping $E_{l-L_m,H} \rightarrow E_{l+1,H}$. Using these setting, we have succeed in implementing the desired polarization-dependent hopping $E_{l-L_m,V} \rightarrow E_{l+1,H}$. Similar discussion also works for the hopping $E_{l+1,H} \rightarrow E_{l-L_m,V}$ with incoming mode $E_{l+1,H}$ from the left BS shown in Fig. 3c. This combined circuit thus realizes the expected hopping Hamiltonian we outlined above.

\section{effective hopping and decay due to the presence of pinhole in the beam splitters}

The presence of pinhole in each beam splitter in this modified cavity circuits distorts the optical modes , which may result in the reduction of effective hopping amplitudes $\kappa$ and other unwanted decay of these modes.
To estimate the influence of the holes, we need to know the explicit distributions of these optical modes inside the cavity. Under the paraxial approximation, the electric fields at two planes $(x_0,y_0,z_0)$ and $(x_1,y_1,z_1)$ can be connected through the Collins integral defined as \cite{collins1970lens}
\begin{flalign}
&e^{-ikz_1}E_1(x_1,y_1) = e^{-ikL}e^{-ikz_0}\frac{i}{\lambda B} \int \int dx_0dy_0 E_0(x_0,y_0) \nn \\
& * \exp \{-\frac{i}{\lambda B} [A(x_0^2+y_0^2)+D(x_1^2+y_1^2)-2(x_0x_1+y_0y_1)], \nn \\ \label{collins}
\end{flalign}
where $\lambda$ and $k$ are the wavelength and wave number of the beams, and $L$ is the length of the optical path between the two planes with the ray transfer matrix defined by $M=\left[ \begin{array}{cc} A & B \\ C & D \end{array}\right]$ \cite{hodgson2005laser}.    Therefore, starting with the Gaussian mode with $l=0$, if we excite the cavity modes such that the beam waists overlap with the SLMs, other cavity modes after passing the SLMs can be obtained by imposing different factors $e^{-il\theta}$ on these beams with different orbital angular momentum $l$. The electric fields at the BSs for different $l$ can then be estimated by the above integral using the transfer matrix $M=\left[ \begin{array}{cc} 0 & f \\ -1/f & 0 \end{array}\right]$ with $f$ the focal length of cavity mirrors.

\begin{figure}[htpb]
    \centering
    \includegraphics[width=.70\linewidth]{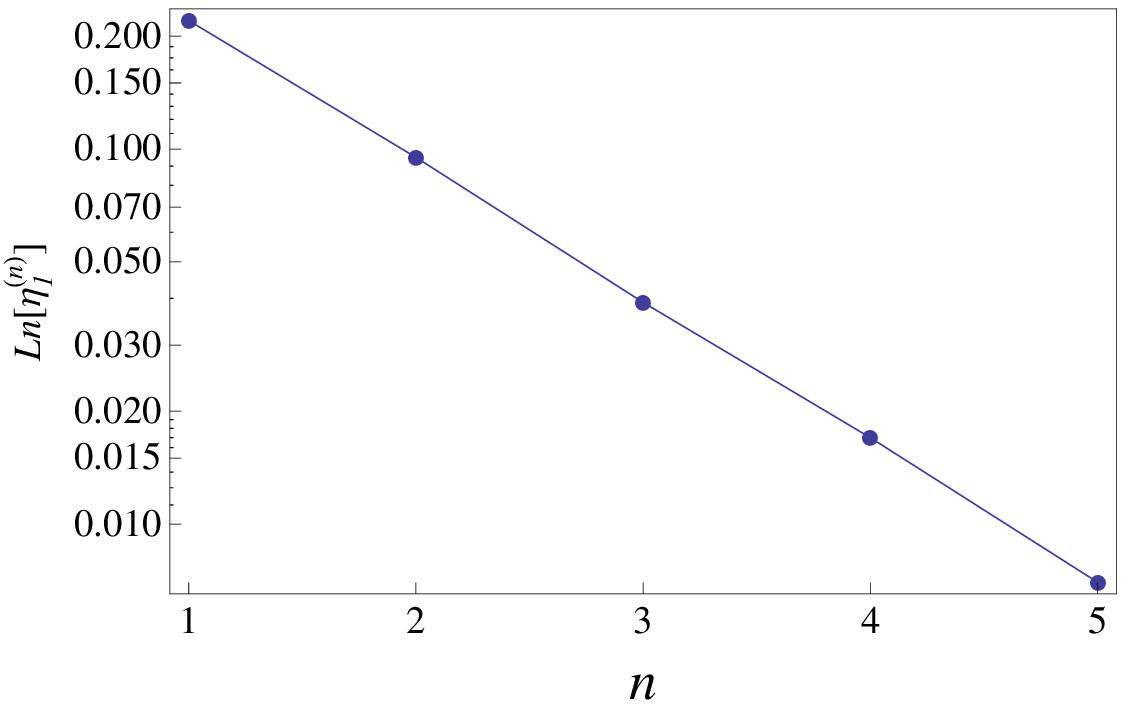}
    \caption{(Color online). Calculated $\eta^{(n)}_1$ for different hopping step $n$, which scales exponentially along with $n$.
     }
    \label{fig:expn}
\end{figure}

Figure 2(b) of the main text shows the calculated optical intensity profiles on the BSs for different $l$. One can see that the $l=0$ mode can be easily distinguished from other modes with $l \neq 0$. To obtain an effective sharp boundary, the hole size is designed such that most portion of $l=0$ mode can go through while most $l=1$ mode is reflected by BSs into the ancillary cavity. This is achieved if we choose the radius $r_h$ of the hole such that $\int_0^{r_h} dr r |E_{l=0}|^2 = \int_{r_h}^{\infty} dr r |E_{l=1}|^2$, where $E_l$ are the electric fields on the BSs obtained from Eq. \ref{collins}.
%Here we neglect the influence of the pinhole on the cavity modes as this effect can be greatly reduced.
For instance, if we choose a large hopping step in the ancillary circuit with $\delta l=\pm(L_m+n)$, then we have $\int_0^{r_h} dr r |E_{l=0}|^2 = \int_{r_h}^{\infty} dr r |E_{l=n}|^2$.
In this case, the logical lattice site $j$ is represented by the composite mode $|l=jn\rangle$.
The influence of the hole on the optical modes $E_n$ can be estimated from
\bea
\eta^{(n)}_1=\int_0^{r_h} dr r |E_n|^2 &\leq & |E_0(r_h)|^{-2}\int_0^{r_h} dr r |E_n|^2|E_0|^2 \nn \\
&\leq & |E_0(r_h)|^{-2}\int_0^{\infty} dr r |E_n|^2|E_0|^2, \nn \\
\eea
which scales as $\eta^{(n)}_1 \sim e^{-n}$ for the usual LG mode $E_l^{p=0}$, and is consistent with our numerical calculation, as shown in Fig. \ref{fig:expn}. Therefore, a larger hopping step is always helpful for the construction of sharp boundaries.

\begin{table}[h]
    \begin{center}
    \begin{tabular}{|c||c|c|c|c|c|}
        \hline
         $n$ & 1 & 2 & 3 & 4 & 5 \\
        \hline
         $r_h(mm)$ & 0.123 & 0.154 & 0.180 & 0.204 & 0.224\\
        \hline
         $\eta^{(n)}_1$ & 0.22 & 0.095 & 0.039 & 0.017 & 0.007\\
        \hline
    \end{tabular}
    \end{center}
    \caption{Calculated radius $r_h$ of the pinhole in the beam splitter and the factor $\eta^{(n)}_1$  for different hopping step $n$. Here we choose the cavity mirrors' focal length $f=10cm$, and the wavenumber $\lambda=0.885\times 10^{-3}mm$. The Gaussian beam for the $l=0$ mode at the SLM is $E_{l=0}=E_0 e^{-r^2/w_0^2}$ with the waist size $w_0=0.2mm$.   }
    \label{table:s1}
\end{table}

%%%--------------------------------------------------------------------------------------
%%%--------------------------------------------------------------------------------------

The reflection of $|l=0\rangle$ mode at the BSs leads to the coupling of the composite modes $|0\rangle$ and $|-n\rangle$, which may soften the boundary we have designed.
This amplitude can be estimated as $\eta^{(n)}_1 \kappa$, as $\eta^{(n)}_1$ describes the portion of photon that has been reflected into the ancillary cavity at BS1 to realize the unwanted hopping $|0 \rangle \rightarrow |l=-n\rangle$.
Meanwhile, the effective coupling between modes $|0\rangle$ and $|n\rangle$ is also weakened, which can be estimated by $(1-\eta^{(n)}_1)\kappa$ since part of the $|l=n\rangle$ modes in the ancillary cavity cannot couple back into the main cavity due to the presence of the pinhole.
Similar discussion also applies to other cavity modes $|l=jn\rangle$, where effective coupling between modes $|(j-1)n\rangle$ and $|jn\rangle$ has been reduced to $(1-\eta^{(n)}_j)\kappa$ with $\eta^{(n)}_j$ defined as $\eta^{(n)}_j=\int_0^{r_h} dr r |E_{jn}|^2$.
The calculation shows that $\eta^{(n)}_j$ also scales exponentially along with the lattice site $j$, as shown in Tab. \ref{table:s2} and Fig. \ref{fig:expj}. This indicates that only the logically adjacent modes with $|l=0\rangle$, and $|l=\pm n\rangle$ are significantly affected due to the presence of the pinhole in the BSs.

\begin{table}[h]
    \begin{center}
    \begin{tabular}{|c||c|c|c|c|c|}
        \hline
         $j$ & 1 & 2 & 3 & 4  \\
        \hline
         $\eta^{(1)}_j$ & 0.22 & 0.036 & 0.004 & $3.0 \times 10^{-4}$ \\
        \hline
         $\eta^{(2)}_j$ & 0.095 & 0.002 & $2.2\times 10^{-5}$ & $1.2 \times 10^{-7}$ \\
        \hline
        $\eta^{(3)}_j$ & 0.039 & $1.3\times 10^{-4}$ & $1.2\times 10^{-7}$ & 0 \\
        \hline
    \end{tabular}
    \end{center}
    \caption{Calculated $\eta^{(n)}_j$ along with the lattice site $j$ for different hopping steps $n=1$, $2$ , and $3$ respectively. Other parameters for cavities are the same with those in table \ref{table:s1}. These parameters have also been token into account during the calculation of results shown in Fig. 4 in the main context. }
    \label{table:s2}
\end{table}

%%%--------------------------------------------------------------------------------------
\begin{figure}[htpb]
    \centering
    \includegraphics[width=.70\linewidth]{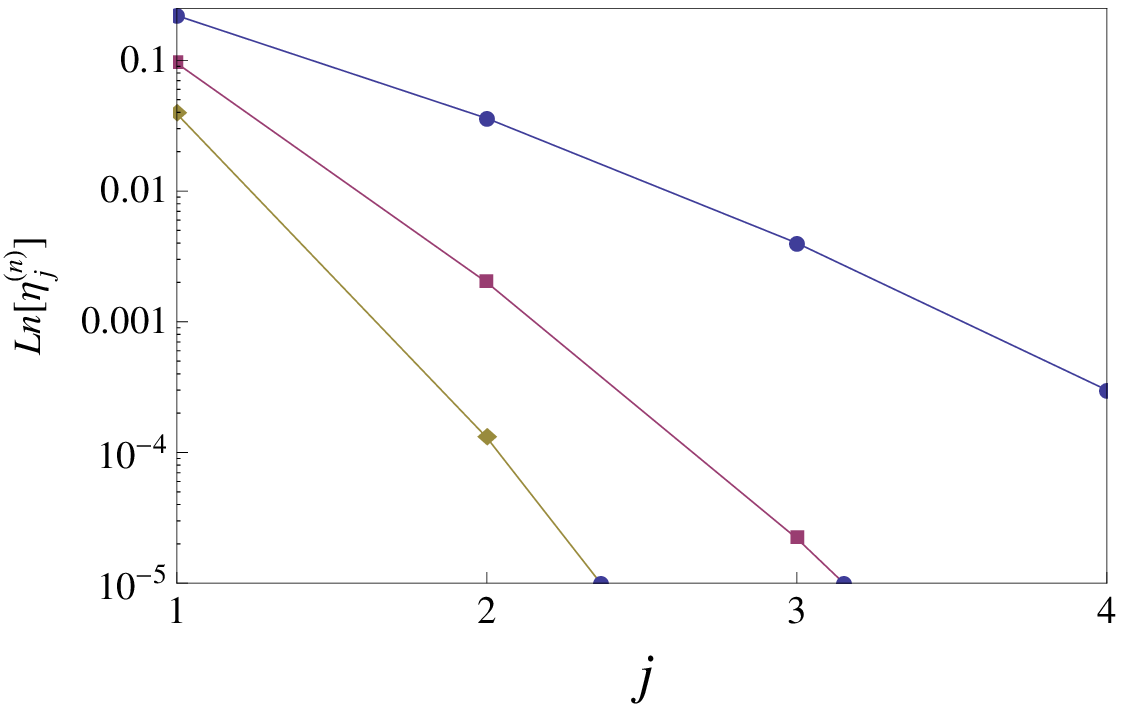}
    \caption{(Color online). Log plots of $\eta^{(n)}_j$ along with the lattice site $j$ for different hopping steps $n=1$, $2$ , and $3$ respectively, which scales exponentially along with $j$.
     }
    \label{fig:expj}
\end{figure}
%%%--------------------------------------------------------------------------------------

Based on the above discussion, the total Hamiltonian of the 1D chain we simulated in Fig. 3 can then be written as
\bea
H_T= H+H'  \label{longH}
\eea
with
\bea
H &=& \sum_{l=0}^{L_m} \big[J_0\cos(\phi) a^{\dag}_{2l} a_{2l+1} +J_1 \alpha^{(n)}_{l+1} a^{\dag}_{2l+1} a_{2l+2} + h.c.\big]\, \nn  \\
H'&=&H_L+H_{L0}+H_{R0}+H_R.  \nn
\eea
Here due to the construction, the system has been divided into three pieces: $H$ is the Hamiltonian of the center piece which we focus on, and takes the exact form of the celebrated SSH model Hamiltonian in the idea case when $\alpha^{(n)}_{l+1}=1$; $H_L$ and $H_R$ represent the Hamiltonian of the left and right subchains and read
\bea
H_L&=&\sum_{l=L_{max}}^{-1} \big[J_1 \alpha^{(n)}_{l} a^{\dag}_{2l-1} a_{2l} + J_0\cos(\phi) a^{\dag}_{2l} a_{2l+1}  + h.c.\big] \nn \\
H_L&=&\sum_{l=L_m+1}^{L_{max}} \big[J_0\cos(\phi) a^{\dag}_{2l} a_{2l+1} +J_1 \alpha^{(n)}_{l+1} a^{\dag}_{2l+1} a_{2l+2} + h.c.\big] \nn
\eea
with  $L_{max}$ the maximal allowed OAM of photonic modes inside the cavity. Finally, the size effect of the pinhole in BSs also induces an additional coupling between $H$ and $H_{L(R)}$ denoted by
\bea
H_{L0}&=&\eta^{(n)}_1J_1 (a^{\dag}_{-1}a_0 +h.c.) \nn \\
H_{R0}&=&\eta^{(n)}_1J_1 (a^{\dag}_{2L_m+1}a_{2L_m+2} +h.c.)  \nn
\eea

In the ideal case $\eta^{(n)}_1 = 0$, the three pieces are totally disconnected. Therefore, we arrive an exact SSH model Hamiltonian as we desire. The presence of finite $\eta^{(n)}_1$ induces weak coupling between the center chain and the bilateral chains, which may soften the boundary effect of the center chain as the edge modes can tunnel into the other two sublattices.  Especially, in the topological nontrivial regime, edge zero-modes appear at the ends of each subchain. In this case, $H_{L0}$ and $H_{R0}$ induce in  effective coupling of these edge modes around each end of the center subchain. This may result in the recombination of the wavefunctions of edge modes and induce a finite energy shift around $\omega=0$.

In Fig. 4,  we have calculated the transmission rate $\tau(\omega)$ for hopping step $n=2$ using parameters we list above. The calculation shows that in topological nontrivial regime with $J'_0/J_1<1$, $\tau(\omega)$ exhibits two different branches around $\omega=0$ with finite energy splitting. This clearly demonstrates the  modification of edge modes due to the presence of weak coupling $H_{L(R)}$ at the ends of the center sublattice. For larger hopping step $n=4$, the splitting of the two branches becomes indistinguishable due to the fast decreasement of $\eta^{(n)}_1$ along with $n$. In this case, we can only focus on the center subchain, while the influence of the bilateral sublattices can be neglected.

%%%--------------------------------------------------------------------------------------
\begin{figure}[htpb]
    \centering
    \includegraphics[width=.970\linewidth]{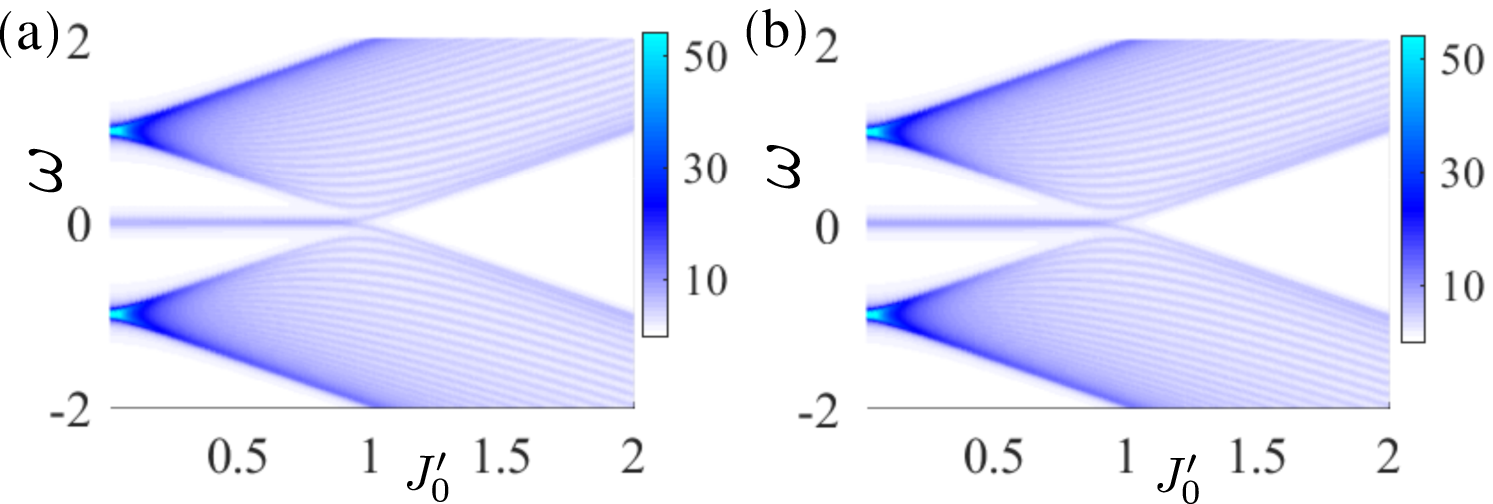}
    \caption{(Color online). (a) and (b) show the calculated transmission rates $\tau(\omega)$ for a prolonged 1D chain with Hamiltonian \ref{longH} and a simplified Hamiltonian $H$ in Eq. (1) respectively.
    The two models almost provide the same $\tau(\omega)$. Other parameters are the same with those in Fig. 4 in the main text.
     }
    \label{fig:expj}
\end{figure}
%%%--------------------------------------------------------------------------------------

We have also calculated $\tau(\omega)$ for 1D chain with soft boundaries. In this case, the presence of edge modes is completely erased, which clearly shows the necessity of boundary construction for observing topological edge modes. Finally, the dynamical properties of edge modes for different hopping steps $n$ are also studied, as shown in Fig. 4(d) and 4(e). When $n=1$, $N_0(t=15)$ exhibits an oscillation along with $J'_0$.  This is due to the interference effect of the two modified edge modes around $\omega=0$ in the presence of finite $\eta^{(n)}_1$. For larger hopping step $n$, such oscillation disappears.

\section{Floquet topological insulator and edge modes inside single cavity}

\begin{figure}[htpb]
    \centering
    \includegraphics[width=0.90\linewidth]{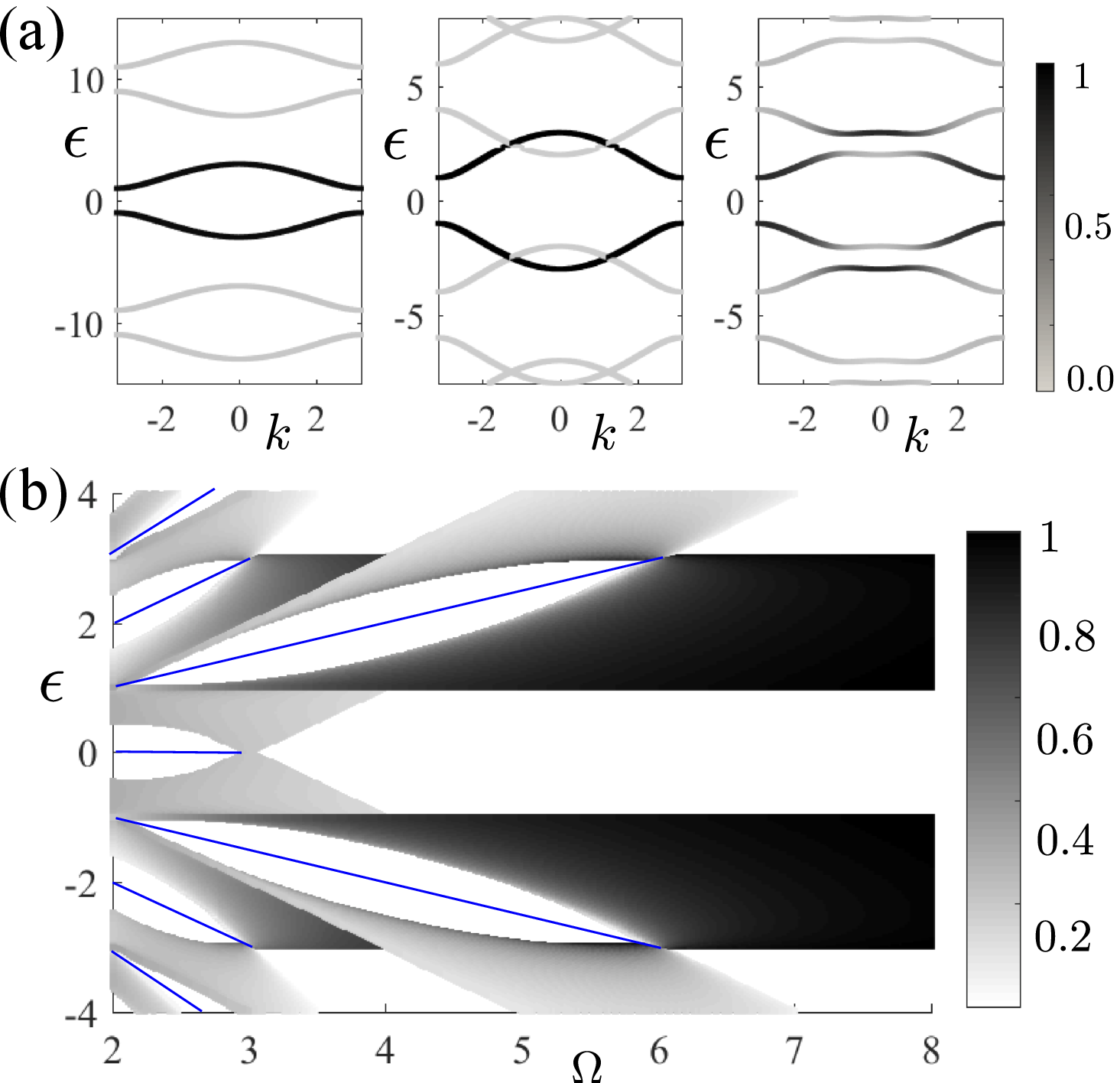}
    \caption{(Color online). Floquet band structures ((a)) and spectrum (b) along with driven frequency $\Omega$ for a topologically trivial static Hamiltonian with $[J_0,J_1,\lambda]=[2,1,1.6]$. The corresponding parameters in the three panels of (a) are $(\Omega,\lambda)=(10,1.6)$,  $(\Omega,\lambda)=(5,0)$, and $(\Omega,\lambda)=(5,1.6)$ respectively (from left to right). The mid-gapped solid lines in (b) indicate the presence of edge states in finite system. The color bar indicates the weight of bulk states on the $m=0$ Floquet subspace.
  }
    \label{fig:s3}
\end{figure}

For given time-periodic system with $H(t)=H(t+T)$, the quasienergy-state can be written as
\bea
\psi(r,t)=e^{-i\epsilon_q t}\phi(r,t)
\eea
where $\epsilon_q$ is the quasienergy and $\phi(r,t)=\phi(r,t+T)$ are periodic functions, which are also named as Floquet modes \cite{levante1995formalized}. These modes satisfy
\bea
H_F(t)\phi_q(t)=\epsilon_q\phi_q(t)
\eea
and
\bea
H_F(t)=H(t)-i\partial_t.
\eea
Since different Floquet modes are physically equivalent when their quasienergies differ by $m\Omega$, it is therefore convenient to consider the first Floquet zone with $\epsilon \in (-\Omega/2, \Omega/2]$. The quasienergies and Floquet modes can be solved in the composite Floquet space $T \otimes R$ where $R$ represents the usual Hilbert space and $T$ is spanned by the periodic functions $\langle t|m\rangle=e^{im\Omega T}$. The index $m$ describes the number of phonons involved in the interaction and defines the subspace named as $m$-th Floquet replica. Therefore in Floquet space, the periodic driven term induce an effective coupling between different Floquet replicas. The corresponding time-independent Floquet Hamiltonian
in Floquet space $T\otimes R$ can be written as
\bea
\mathbb{H}_F=\hat{F}_m\otimes\hat{H}^{(m)}+\Omega \hat{F}_z\otimes I_R,
\eea
where $I_R$ means the identity operator in usual Hilbert space, $H(t)=\sum_m \hat{H}^{(m)}e^{im\Omega t}$, $(F_m)_{i,j}=\delta_{j,i+m}$, and $(F_z)_{m,n}=m\Omega\delta_{m,n}$.
Using this formalism, the wave function in usual Hilbert space $|\psi\rangle=\sum_{m,j}c_{m,j}(t)\exp(im\Omega t)|j\rangle$ can be rewritten as $|\psi^F\rangle=\sum_{m,j}c_{m,j}(t)|m,j\rangle$, which satisfies the so-called Floquet Schr\"{o}dinger equation
\bea
\frac{d|\psi^F\rangle}{dt}=-i \mathbb{H}_F |\psi^F\rangle - \frac{\gamma}{2}|\psi^F\rangle
\eea
by taking into account the dissipation effect. This leads to the following evolution equations for each coefficient as
\bea
\frac{d c_{m,j}(t)}{dt}=-i [\mathbb{H}_F]_{mj,m'j'} c_{m',j'}(t) - \frac{\gamma_j}{2}c_{m,j}(t).
\eea
The formal solution in frequency domain can be written as
\bea
c_{m,j}(\omega)=\frac{-1}{\sqrt{2\pi}} \langle m, j| \frac{1}{\omega+i\frac{\Gamma^F}{2}-\mathbb{H}_F}|\psi^F(0)\rangle
\eea
with $|\psi^F(0)\rangle$ representing the initial state vector and $\Gamma^F=F_0\otimes \Gamma$ the decay matrix in Floquet space. Mathematically, if we sum over all intermediate state $|m,j\rangle$ together with all possible initial vector $|\psi^F(0)\rangle$, we obtain the following total output spectrum as
\bea
T(\omega)=\sum_{\psi^F(0)}\sum_{m,j} |c_{m,j}(\omega)|^2   \nn
\eea
%where we have assumed $\mathbb{H}_F=\sum_{D_F}D_F|D_F\rangle\langle D_F |$ and $A_{D_F,m,j}(\omega)=\langle m, j|(\omega+i\gamma_j/2-D_F)^{-1}|D_F\rangle$ with $|D_F\rangle$ the single-particle eigenvectors of Floquet Hamiltonian with the eigenvalues $D_F$.
Fig. \ref{fig:s3} shows the quasienergy spectra as a function of driven frequency $\Omega$. We note that although the original static Hamiltonian is topological trivial for given parameters, a topological transition occurs when different replicas start to overlap with each other.
In addition, the interaction of different Floquet replicas usually leads to the appearance of new gaps at $ m \Omega/2$ with $m \in Z$ when lowering $\Omega$.

To show how $T(\omega)$ can be distilled from the outputs of cavity modes, we rewrite the wavefunction in the usual Hilbert space as $|\psi\rangle=\sum_{j}d_j(t)|j\rangle$ with $d_j(t)=\sum_m c_{m,j}(t)\exp(im\Omega t)$.
Since $c_{m,j}(t)$ changes much slowly during a single time period, the spectrum $c_{m,j}(\omega)$ will mainly distributes within the regime $\omega \in (-\Omega/2,\Omega/2)$. Therefore, for given frequency $\omega$, we can approximate $d_j(\omega)\simeq c_{M,j}(\omega-M\Omega)$ with $M=[\omega/\Omega]$ the integer closest to $\omega/\Omega$. Finally, we have
\bea
T(\omega) &=& \sum_{\vec{\psi}(0)}\sum_{m,j} c^*_{m,j}(\omega)c_{m,j}(\omega) \nn  \\
 &=& \sum_{\vec{\psi}(0)}\sum_{m,j} d^*_{j}(\omega+m\Omega)d_{j}(\omega+m\Omega).  \label{Tw}
\eea
By monitoring the dynamics of different modes inside the cavity, we can obtain the output spectrum for $\omega \in (-\Omega/2,\Omega/2)$. In addition, when $\omega$ is resonant with Floquet modes, it will produce a peak in the spectrum. This provides an effective method to detect various Floquet gaps and edge modes in our system.

One of the particular properties of periodic driven system is presence of Floquet topological transition and mid-gapped edge modes even when their static Hamiltonian is topologically trivial. These features can be manifested by detecting the output spectrum $T(\omega)$ as outlined above. Compared with its static counterpart, Floquet topological system have two different kind of topological edge modes at $\epsilon=0$ and $\Omega/2$.  The presence of finite gap around $\epsilon=0$ and $\Omega/2$ also enable us to detect Floquet topological phase transitions by observing the output spectrum due to the changes of edge modes. Specifically, when $\epsilon=0$ or $\Omega/2$, mid-gapped edge modes contributes mostly to the summation in \ref{Tw}, while all contributions from other Floquet modes are greatly reduced due to the presence of finite gaps. The amplitude of $T(\omega)$ is approximately proportional to the total number of edge modes, which can be viewed as a direct evidence of Floquet topological phase transitions.

%%%%--------------------------------------------------------------------------------------
%%%%--------------------------------------------------------------------------------------

Floquet topological system usual exhibits distinct topological features compared with their static ones. The presence of different edge modes indicates that we need new topological invariants. The single-particle Bloch Hamiltonian reads
\bea
H(t,k)=[B_x(k)+  \lambda\cos(\Omega t)] \sigma_x + B_y(k) \sigma_y.
\eea
Since the system possesses chiral symmetry as $\sigma_z H(t,k) \sigma_z= -H(-t,k)$, the topological invariants associated with the number of end states $v_0$ and $v_+$ at $\epsilon=0$ and $\Omega/2$ can be obtained from their time evolution operator defined as \cite{asboth2014chiral,dal2015floquet}
\bea
F(k)= \mathbb{T} \exp[-i\int_0^{T/2}H(t,k)]
\eea
with $\mathbb{T}$ the time-ordering operator. Thanks to the chiral symmetry, the evolution of the second part of the cycle can be written as
\bea
\sigma_z F(k) \sigma_z= \mathbb{T} \exp[-i\int_{T/2}^{T}H(t,k)]
\eea
If we express $F(k)$ in the canonical form as
\bea
F(k)=\left(
\begin{array}{cc} a(k) & b(k) \\ c(k) & d(k) \end{array}
\right),
\eea
the winding number of blocks $b(k)$ and $d(k)$ gives the number of edge states at quasienergy $\epsilon=0$ and $\Omega/2$ respectively
\bea
v_0=\frac{1}{2\pi i} \int_{-\pi}^{\pi} dk \frac{d}{dk} \ln \det b(k), \nn \\
v_+=\frac{1}{2\pi i} \int_{-\pi}^{\pi} dk \frac{d}{dk} \ln \det d(k).  \nn
\eea
We note that the above invariants are well defined only when the gap at $\epsilon=0$ and $\Omega/2$ are not close. When the gaps are closed, $b(k)$ or $d(k)$ may has an eigenvalues of zero, which invalids the above definition.

\bibliographystyle{apsrev4-1}

\bibliography{tex-total}